\documentclass[11pt]{article}
\usepackage{times}
\usepackage{sidecap}
\usepackage{geometry}
\usepackage{floatrow}
\usepackage{float}
\usepackage{hyperref}
\usepackage{graphics,graphicx}
\usepackage{subcaption}
\usepackage{epsf,wrapfig}

\usepackage{natbib,bm}

\usepackage{paralist}
\let\olditem\item
\renewenvironment{thebibliography}[1]{%
  \section*{\refname}
  \let\par\relax
  \renewcommand{\item}[1][]{\olditem[\textbullet]}%
  \inparaenum}{\endinparaenum}

\newcounter{aff}

\DeclareGraphicsExtensions{.jpg,.pdf,.png,.eps}

\usepackage{caption}
\usepackage{xcolor}

\usepackage[most]{tcolorbox}

\bibpunct{(}{)}{;}{a}{}{,}

\usepackage{soul}

\definecolor{darkcyan}{rgb}{0.0, 0.0, 0.7}

\definecolor{boxcolor}{RGB}{193, 229, 255}
\definecolor{figboxcolor}{RGB}{193, 229, 255}
\definecolor{boxcolorgray}{RGB}{227, 227, 227}
\definecolor{darkcyan}{rgb}{0.0, 0.0, 0.7}

\sethlcolor{boxcolorgray}

\def\gtrsim{\lower 2pt \hbox{$\, \buildrel {\scriptstyle >}\over
{\scriptstyle \sim}\,$}}
\def\lesssim{\lower 2pt \hbox{$\, \buildrel {\scriptstyle <}\over
{\scriptstyle \sim}\,$}}

\def\approxlt{\lower.2em\hbox{$\buildrel < \over \sim$}}
\def\approxgt{\lower.2em\hbox{$\buildrel > \over \sim$}}

\def\be{\begin{equation}}
\def\ee{\end{equation}}
\def\bea{\begin{eqnarray*}}
\def\eea{\end{eqnarray*}}

\def\jref@jnl#1{{\rm#1}}
\def\aj{\jref@jnl{AJ}}			
\def\araa{\jref@jnl{ARA\&A}}		
\def\apj{\jref@jnl{ApJ}}			
\def\apjl{\jref@jnl{ApJ}}		
\def\apjs{\jref@jnl{ApJS}}		
\def\ao{\jref@jnl{Appl.~Opt.}}		
\def\apss{\jref@jnl{Ap\&SS}}		
\def\aap{\jref@jnl{A\&A}}		
\def\aapr{\jref@jnl{A\&A~Rev.}}		
\def\aaps{\jref@jnl{A\&AS}}		
\def\azh{\jref@jnl{AZh}}			
\def\baas{\jref@jnl{BAAS}}		
\def\jrasc{\jref@jnl{JRASC}}		
\def\memras{\jref@jnl{MmRAS}}		
\def\mnras{\jref@jnl{MNRAS}}
\def\nar{\jref@jnl{NewAR}}
\def\pra{\jref@jnl{Phys.~Rev.~A}}	
\def\prb{\jref@jnl{Phys.~Rev.~B}}	
\def\prc{\jref@jnl{Phys.~Rev.~C}}	
\def\prd{\jref@jnl{Phys.~Rev.~D}}	
\def\pre{\jref@jnl{Phys.~Rev.~E}}	
\def\prl{\jref@jnl{Phys.~Rev.~Lett.}}	
\def\pasp{\jref@jnl{PASP}}		
\def\pasj{\jref@jnl{PASJ}}		
\def\qjras{\jref@jnl{QJRAS}}		
\def\skytel{\jref@jnl{S\&T}}		
\def\solphys{\jref@jnl{Sol.~Phys.}}	
\def\sovast{\jref@jnl{Soviet~Ast.}}	
\def\ssr{\jref@jnl{Space~Sci.~Rev.}}	
\def\zap{\jref@jnl{ZAp}}			
\def\nat{\jref@jnl{Nature}}		
\def\iaucirc{\jref@jnl{IAU~Circ.}}
\def\aplett{\jref@jnl{Astrophys.~Lett.}}
\def\apspr{\jref@jnl{Astrophys.~Space~Phys.~Res.}}
\def\bain{\jref@jnl{Bull.~Astron.~Inst.~Netherlands}}
\def\fcp{\jref@jnl{Fund.~Cosmic~Phys.}}
\def\gca{\jref@jnl{Geochim.~Cosmochim.~Acta}}
\def\grl{\jref@jnl{Geophys.~Res.~Lett.}}
\def\jcp{\jref@jnl{J.~Chem.~Phys.}}	
\def\jgr{\jref@jnl{J.~Geophys.~Res.}}	
\def\jqsrt{\jref@jnl{J.~Quant.~Spec.~Radiat.~Transf.}}
\def\memsai{\jref@jnl{Mem.~Soc.~Astron.~Italiana}}
\def\nphysa{\jref@jnl{Nucl.~Phys.~A}}
\def\physrep{\jref@jnl{Phys.~Rep.}}
\def\physscr{\jref@jnl{Phys.~Scr}}
\def\planss{\jref@jnl{Planet.~Space~Sci.}}
\def\procspie{\jref@jnl{Proc.~SPIE}}

\def\ion#1#2{#1$\;${\small\rm\@Roman{#2}}\relax}

\newcount\lecurrentfam
\def\LaTeX{\lecurrentfam=\the\fam \leavevmode L\raise.42ex
\hbox{$\fam\lecurrentfam\scriptstyle\kern-.3em A$}\kern-.15em\TeX}
\def\sizrpt{
(\fontname\the\font): em=\the\fontdimen6\font, ex=\the\fontdimen5\font
\typeout{
(\fontname\the\font): em=\the\fontdimen6\font, ex=\the\fontdimen5\font
}}
\definecolor{dark}{rgb}{ 0.34, 0.09, 0.49}

\usepackage{fancyhdr,url}
\usepackage{transparent}
\usepackage{tablefootnote}
\usepackage{mathpazo}

\begin{document}

% Title and subtitle
\noindent
\begin{raggedright}
{\Huge\bfseries
mm-jets in the multi-messenger era \\[0.5em]
\Large The launching, collimation, acceleration, feedback and evolution of astrophysical jets across all scales
}\end{raggedright}\\

%\vfill
% Author block
%\thispagestyle{empty}
%\vspace{1em}

\noindent Karri I. I. Koljonen$^{1,2}$, 
Thomas D. Russell$^{3}$, 
Britton Jeter$^{4,2}$,
Lauren Rhodes$^{5}$,
Alexandra J. Tetarenko$^{6}$,
Mar\'ia D\'iaz Trigo$^{7}$,
Talvikki Hovatta$^{4,2}$,
Pau Bosch-Cabot$^{6}$,  
Mamta Pandey-Pommier$^{8}$, 
Alessio Marino$^{9}$,
Jakob van den Eijnden$^{10}$, 
Dieu D. Nguyen$^{11}$,
Melania Del Santo$^{3}$,
Thomas Maccarone$^{12}$,
Anna Y. Q. Ho$^{13}$,
Maria C. Baglio$^{14}$, 
Sera Markoff$^{10,15}$, 
and the ALMA2040 Transient and Time-Domain Astronomy working group

\vspace{1em}
{
\footnotesize{
\noindent $^{1}$ Institut for Fysikk, Norwegian University of Science and Technology, H{\o}gskloreringen 5, Trondheim, 7491, Norway \\
$^{2}$ Finnish Centre for Astronomy with ESO (FINCA), FI-20014 University of Turku, Finland \\
$^{3}$ INAF, Istituto di Astrofisica Spaziale e Fisica Cosmica, via U. La Malfa 153, I-90146 Palermo, Italy  \\
$^{4}$ Aalto University Metsähovi Radio Observatory, Metsähovintie 114, 02540 Kylmälä, Finland \\
$^{5}$ Department of Physics, McGill University, 3600 Rue University, Montréal, QC H3A 2T8, Canada \\
$^{6}$ Department of Physics and Astronomy, University of Lethbridge, Lethbridge, Alberta T1K 3M4, Canada \\
$^{7}$ European Southern Observatory (ESO), Karl-Schwarzschild-Strasse 2, D-85748 Garching bei München, Germany  \\
$^{8}$ Pole Scientific, University Catholic of Lyon- University of Lyon, 10 place des Archives 69002, Lyon, France \\
$^{9}$ Departamento de Física, Universidad de Santiago de Chile (USACH), Av. Víctor Jara 3493, Estación Central, Chile \\
$^{10}$ Anton Pannekoek Institute for Astronomy, Universiteit van Amsterdam, Science Park 904, NL-1098 XH, Amsterdam, the Netherlands \\
$^{11}$ Department of Astronomy, University of Michigan, 1085 South University Avenue, Ann Arbor, MI 48109, USA\\
$^{12}$ Department of Physics \& Astronomy, Texas Tech University, Box 41051, Lubbock, TX, 79409-1051, USA \\
$^{13}$ Department of Astronomy, Cornell University, Ithaca, NY 14853, USA \\
$^{14}$ INAF–Osservatorio Astronomico di Brera, INAF, Via Bianchi 46, Merate, I-23807, Italy \\
$^{15}$ Institute of Astronomy, University of Cambridge, Madingley Road, Cambridge CB3 0HA, UK \\
}
}
\begin{tcolorbox}[boxsep=3pt,left=0pt,right=0pt,top=1pt,bottom=1pt,colframe=boxcolor,colback=boxcolor]

\justify
\textbf{Abstract:} Jets and outflows operate across the full cosmic mass scale from young stellar objects (YSOs) and X-ray binaries (XRBs) to active galactic nuclei (AGN), gamma-ray bursts (GRBs), and tidal disruption events (TDEs), shaping many time-domain and multi-messenger transients. Key questions concern on how jets are launched, accelerated and collimated; where and how they dissipate energy and accelerate particles; and how they couple to their environments to set electromagnetic counterpart diversity and feedback. Millimeter and sub-millimeter observations uniquely address these questions by tracking the evolving synchrotron turnover for robust calorimetry and Lorentz-factor constraints, providing full-Stokes polarization and Faraday diagnostics close to the jet base, and enabling high-resolution imaging from Galactic ejecta to event-horizon-scale AGN jet bases via global Very Long Baseline Interferometry (VLBI). We describe the jet and outflow science enabled by ALMA2040, a next-generation mm/sub-mm interferometer combining an order-of-magnitude sensitivity gain with rapid-response observing, simultaneous multi-band coverage, full-Stokes polarimetry, and integration with global VLBI networks. These capabilities will enable statistically controlled, trigger-driven samples spanning XRB outbursts, early GRB afterglows, protostellar disk winds and jet launching, AGN jet bases, and jetted TDEs, establishing mm/sub-mm astronomy as a cornerstone of time-domain and multi-messenger jet physics in the 2040s and beyond.
\end{tcolorbox}

\newpage{}

\vspace{-0.7cm}
\section{Scientific Context and Open Questions}

In the 2040s, time-domain and multi-messenger astronomy will be routinely driven by external triggers. Third-generation gravitational wave (GW) observatories will localize compact-object mergers, next-generation neutrino and very-high-energy (VHE) gamma-ray facilities will deliver time-critical alerts, and wide-field optical/X-ray surveys will provide a steady flow of transient discoveries. In many of these systems, \emph{relativistic jets} set the observable outcome: they power GRBs, shape the electromagnetic counterparts of the subset of mergers that launch outflows, accompany a fraction of TDEs, and couple accretion to feedback in AGN and XRBs. 

In this context, mm/sub-mm observations provide a unique and often missing set of diagnostics. They capture the rapidly evolving synchrotron spectral turnover needed for robust calorimetry (jet energy, Lorentz factor and geometry), provide full-Stokes polarization that probes magnetic field ordering, magnetization and Faraday effects, and deliver the angular resolution needed to image jet launching regions in nearby AGN and track ejecta in Galactic sources. ALMA2040 is designed to deliver these capabilities with the sensitivity, cadence and polarimetric capabilities required for \emph{statistically controlled, trigger-driven samples}---turning today’s heterogeneous case studies into population-level constraints through targeted follow-up of larger and more representative event sets (including fainter, more distant and off-axis jets).

\begin{tcolorbox}[title=ALMA2040 will directly address three central open questions:, colback=blue!5, colframe=blue!60!black]
\vspace{0.1cm} \noindent \textbf{(1) How are jets launched, accelerated and collimated?}
What sets the jet Lorentz factor, opening angle and magnetization across black holes and neutron stars?

\vspace{0.1cm} \noindent \textbf{(2) Where and how are particles accelerated in jets?}
Are dissipation and high-energy emission powered primarily by shocks or magnetic reconnection, and how is this linked to polarization and spectral evolution?

\vspace{0.1cm} \noindent \textbf{(3) How do jets couple to their environments and connect to multi-messenger counterparts?}
How and where is energy deposited (from AU to kpc scales), and how does this determine the diversity of electromagnetic counterparts to mergers, GRBs and nuclear transients?
\end{tcolorbox}

\vspace{0.1cm} \noindent \textbf{Jet formation and evolution across mass and timescales.}
In every jet producing system, the key missing step is linking the state of the inflow (accretion rate, geometry and magnetic flux) to measurable jet properties (Lorentz factor, opening angle, magnetization and stratification) on the timescales on which the jet is evolving. XRBs provide controlled experiments where these quantities vary on hours to years timescales during outbursts \citep[e.g.,][]{Corbel2013, Tetarenko2016, tetarenko21}, while AGN provide access to horizon-scale launching regions through variability, polarization and mm/sub-mm VLBI \citep[e.g.,][]{Yuan2014, Blandford2019}. 

In YSOs, outflow launching likewise remains difficult to resolve spatially and temporally because the relevant scales are at sub-au radii and the flow is highly variable \citep[e.g.,][]{Frank2014, Bally2016, Bjerkeli2016}. At the most extreme end, GRBs, merger-driven relativistic outflows, and TDEs provide complementary probes of jet formation in catastrophic events \citep[e.g.,][]{Piran2004, Kumar2015, Metzger2020, Alexander2020}. GRBs launch the fastest known astrophysical jets, yet reverse shock emission is rarely captured because it peaks on minute-to-hour timescales and often near mm/sub-mm frequencies, requiring rapid response and simultaneous multi-band coverage \citep[e.g.,][]{Sari1999,Nakar2004,laskar19}. While most TDEs lack bright early-time radio emission, a small fraction launch relativistic jets whose mm emission at early times constrains Lorentz factors, energetics, and circumnuclear environments \citep[e.g.,][]{Zauderer2011,Metzger2012,Zauderer2013,Eftekhari2018}. For a broader discussion of jet-driven \emph{cosmic explosions} including merger afterglows and other fast transients, see the companion ALMA2040 white paper \emph{Understanding cosmic explosions}.

\vspace{0.1cm} \noindent \textbf{Jet composition, structure and magnetic fields.}
A central goal is to measure jet magnetization and magnetic field ordering at the emission site and to connect these to dissipation (flares), particle acceleration and high-energy messengers requiring high-S/N multi-band polarization capabilities \citep[e.g.,][]{laskar19,2019ApJ...884L..58U,Lister2021}. Determining how magnetization regulates ultra-relativistic outflows and particle acceleration requires joint constraints from spectra, polarization and variability \citep[e.g.,][]{Tchekhovskoy2011,Sironi2014,Kumar2015}. XRBs and AGN show rapid stochastic variability in the mm band that traces the structure of their inner jets \citep[e.g.,][]{tetarenko21,zdziarski22,Hovatta2009}. A key open problem is to identify the dissipation sites and conditions that produce both mm-band variability and high-energy messengers \citep[e.g.,][]{IceCube_2018,Keivani2018,Kouch2024}; this requires rapid, contemporaneous measurements of spectral-turnover evolution and full-Stokes polarization during gamma-ray and neutrino alerts.

\vspace{0.1cm} \noindent \textbf{Jet–environment interaction.} Jet feedback is ultimately constrained by calorimetry: measuring how kinetic power is transferred to ambient gas through shocks, bubbles, lobes and entrainment, and where non-thermal particles are accelerated. This demands sensitive imaging and spectroscopy of interfaces between jet and the interstellar medium (ISM) and robust broadband spectra to separate aged from freshly accelerated particle populations. At larger scales, jet--ISM interfaces in both XRBs and AGN mark sites of non-thermal particle acceleration that may reach TeV--PeV energies, though the microphysics of these shocks remains uncertain \citep[including the amplification of magnetic fields and the generation of cosmic rays; ][]{Caprioli2014a,Caprioli2014b}. In XRBs, and in jet-inflated bubbles around ultra-luminous XRBs (ULXs), resolved nebulae and hotspots provide calorimetry of time-averaged jet power and outflow energetics \citep[e.g.,][]{Gallo2005,Cseh2012,Soria2014}. In AGN, shocks, lobes, and X-ray cavities inflated by jets deposit energy into their surroundings and provide laboratories for particle acceleration and plasma physics on galactic and cluster scales \citep[e.g.,][]{McNamara2007,Hardcastle2020}. These interactions constrain fundamental jet properties, including kinetic power, speed and composition, through lobe and nebula energetics, bow-shock dynamics and spectral ageing.

\section{Science Case for the 2040's}

\begin{figure}[t!]
\begin{center}
\includegraphics[height=8.5cm]{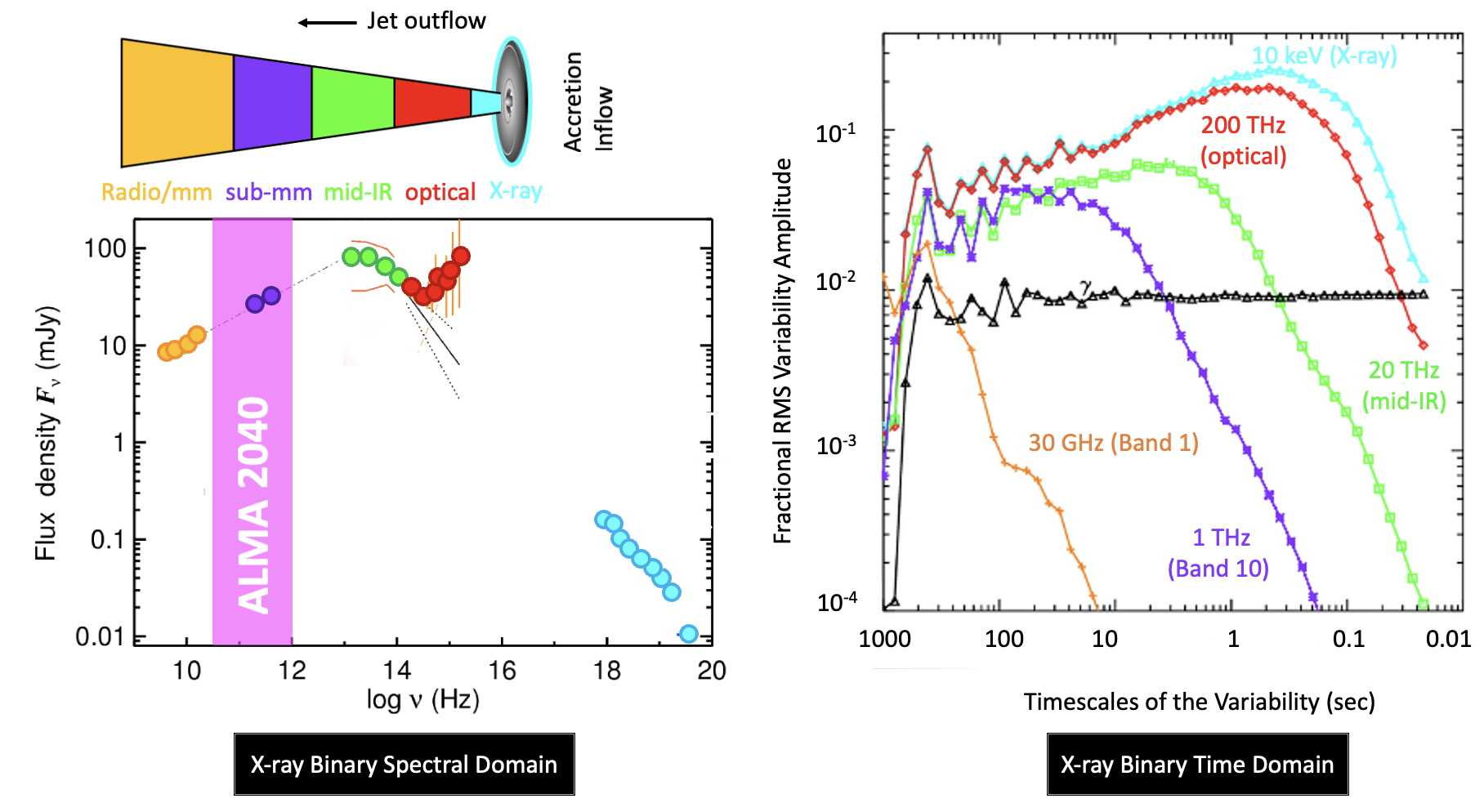}
\end{center}
\caption{\label{fig:xrb} Illustration of the XRB spectral-timing phase space \citep{gan11}. {\it Left:} A schematic of emission regions ({\it top}) and a snapshot of a characteristic XRB broad-band spectrum ({\it bottom}). {\it Right:} Fourier power spectra (from a jet-based theoretical model, where the black curve represents the matter fluctuations injected into the jet; \citealt{mal14}), showing that shorter wavelengths probe closer to the compact object and exhibit higher-amplitude variability on shorter timescales. The ALMA2040 bands fill an order-of-magnitude gap in wavelength coverage and are expected to show strong variability on $\gtrsim$0.1\,s timescales (purple curve).} 
\end{figure} 

\noindent \textbf{XRBs: real-time jet evolution, feedback and multi-messenger links.\,\,} In XRBs, mm-band monitoring has revealed that jet emission can evolve on seconds‑to‑days timescales, enabling direct tests of particle acceleration, shock formation, and jet--ISM coupling \citep{tetarenko21,zdziarski22}. Jet feedback occurs both through persistent or repeatedly launched jets that inflate long-lived structures and drive localized energy injection into the ISM \citep{tetarenko18,tetarenko20}, and through transient ejecta whose interactions can be traced via faint continuum and sensitive molecular line diagnostics \citep{bosch2026constraining}. Millimeter detections from neutron star binaries have opened new avenues for testing outflow mechanisms in strong magnetic fields and isolating the role of compact object properties in launching and dissipating outflows \citep{vandeneijnden22,baglio23,russell24,koljonen25}. Jets in neutron star XRBs can brighten in the radio following thermonuclear bursts, enabling jet-speed estimates \citep{russell24}; extending this approach to the mm band should improve constraints because variability is faster and optical depth is lower. XRBs are also emerging as multi-messenger targets, with ultra-high-energy (UHE) gamma-ray detections and tentative neutrino associations motivating rapid, broadband follow-up during high-energy activity \citep[e.g.,][]{lhaaso24,koljonen23,neronov25}.

ALMA2040’s sensitivity, bandwidth, cadence and polarimetric capabilities will enable time-resolved measurements of jet spectral breaks and power-spectral features across the radio--mm continuum, constraining jet stratification/geometry, bulk speed, opening angle, magnetic field strength near the jet base and the magnetic flux threading the compact object (Figure~\ref{fig:xrb}; \citealt{tetarenko21,zdziarski22}). High-cadence mm polarimetry, especially when combined with timing, can disentangle emission components and track magnetic field ordering during flares, testing shock- and reconnection-driven dissipation scenarios. Increased sensitivity and long baselines will enable tracking of ejecta and jet--ISM interaction \citep[e.g.,][]{Gallo2005,tetarenko18} and will extend these measurements to radio-faint systems \citep[including many neutron star XRBs, e.g.,][]{Migliari2010,koljonen25} and to more distant populations in the outer Galaxy, the Magellanic Clouds and nearby galaxies \citep[e.g.,][]{CrookMansour2026}. Unusually high radio/mm-to-X-ray flux ratios in external galaxies can flag Wolf--Rayet XRBs, which may be bright radio/mm and GeV--PeV $\gamma$-ray emitters powered by jet interactions with dense stellar winds \citep[e.g.,][]{Tavani2009,Dubus2010,Koljonen2018,LHAASO_2025,Zdziarski2026} and potential progenitors of double black holes \citep[e.g.,][]{Belczynski2013}.

\begin{figure}
    \centering
    \includegraphics[width=0.5\linewidth]{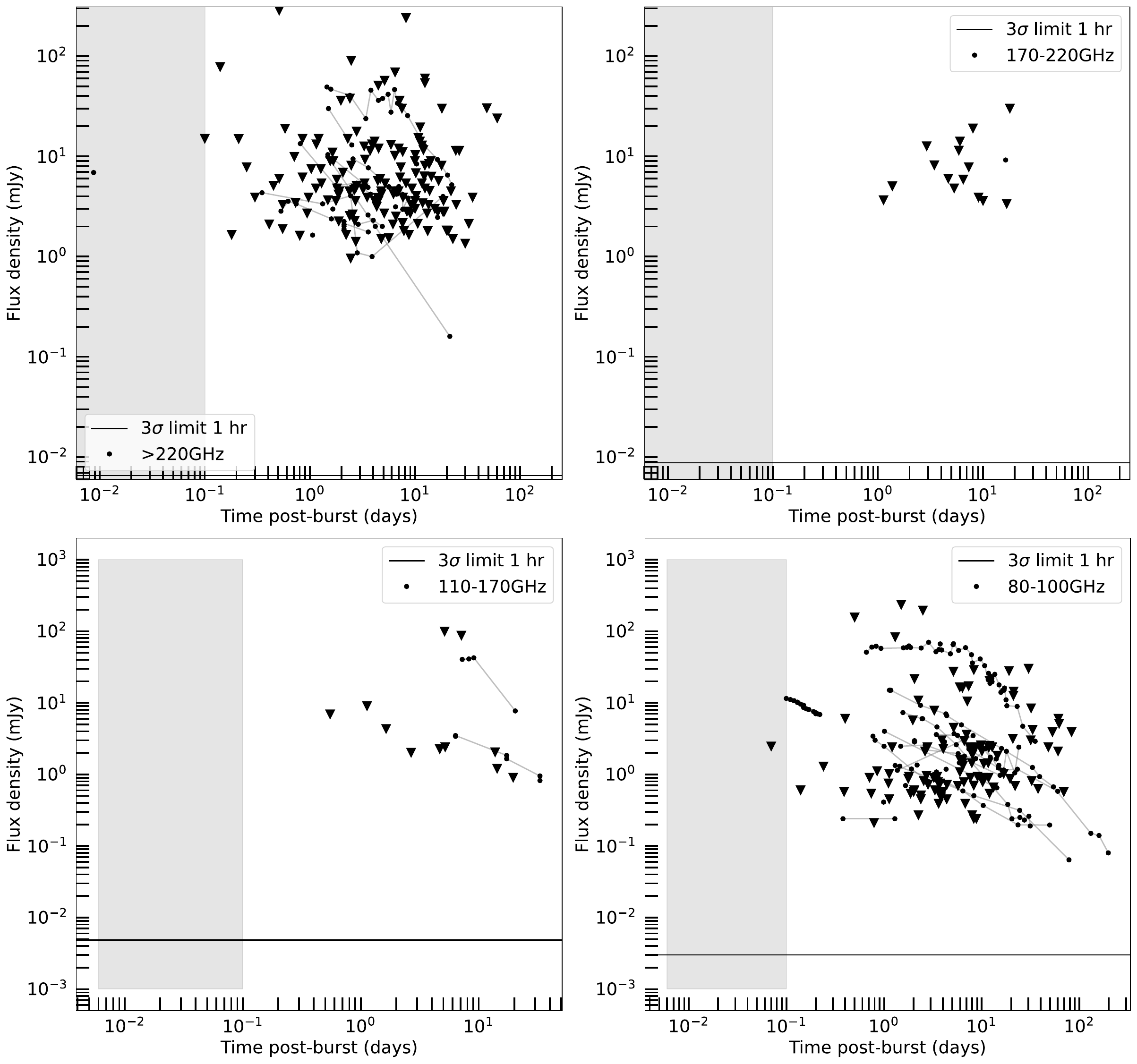}
    \caption{Sub-mm observations of long GRBs with current generation of sub-mm telescopes \citep[][and references therein]{2012A&A...538A..44D}. The horizontal black lines show representative ALMA2040 1-hr continuum sensitivities. The gray region indicates the expected reverse-shock-dominated window at each frequency \citet{bright2023}. Only a single detection has been made on these timescales with a rapid-response mode \citep{2026ApJ..1005L..31K}. }
    \label{fig:GRBs_submm}
\end{figure}

\vspace{0.1cm} \noindent \textbf{GRBs: capturing the reverse shock and afterglow onset.\,\,} Millimeter observations have transformed GRB studies by capturing the brief early afterglow that encodes jet magnetization, Lorentz factors and shocked ejecta structure \cite[][]{zhang05,granot02,gao13, 2021Galax...9...82G, 2021ApJ...913...58S}. Polarized mm emission constrains magnetic topology \cite{laskar19, 2019ApJ...884L..58U}, yet most GRBs fade before current facilities can respond. Proof-of-concept demonstrations already show that minute-scale sub-mm Target-of-Opportunity (ToO) response is operationally feasible; e.g., \citet{2026ApJ..1005L..31K} report an SMA detection obtained within $\sim$12.6 minutes post-burst, directly accessing the reverse-shock/afterglow-onset window.

With an order-of-magnitude sensitivity gain and rapid-response observing, ALMA2040 will deliver early-time multi-band detections for a large fraction of GRBs that would be non-detections today (Figure~\ref{fig:GRBs_submm}), enabling systematic measurements of afterglow onset and reverse shock emission on 10--60 minute timescales, yielding robust measurements of jet Lorentz factors \cite[][at cm-wavelengths]{bright2023} and magnetization \cite{2018ApJ...862...94L}. Simultaneous multi-band coverage is essential during this fast evolving phase, and coordination with upcoming high-energy, GW and neutrino observatories will fully integrate GRBs into the multi-messenger framework. The same rapid-response, multi-band mm/sub-mm capability is critical for nearby GW-detected mergers with jets, where the spectral peak provides clean calorimetry and strong constraints on jet angular structure and viewing angle (see also the companion ALMA2040 white paper \emph{Understanding cosmic explosions}). ALMA2040 will therefore turn early mm/sub-mm afterglows from rare detections into routine constraints on the Lorentz factor, magnetization and ejecta structure on the required reverse shock timescales.

\vspace{0.1cm} \noindent \textbf{YSOs: resolving disk winds, jet launching and magnetic fields.\,\,} Jets and disk winds are ubiquitous accretion phenomena in YSOs and provide the nearest laboratory for studying magnetized outflows \cite{Frank2014}. ALMA has resolved the bases of collimated jets and molecular disk winds, revealing stratified structures that include high velocity jets, rotating molecular disk winds, shocked winds and entrained outflow shells \citep{Lee2017,Bjerkeli2016,Tabone2020,LopezVazquez2024}. Spectral‑line observations of \textsc{CO}, SiO, SO, [\textsc{CI}], and other molecular tracers probe kinematics, excitation and chemistry of these components, enabling measurements of launching radii, angular momentum extraction, mass-loss rates and jet interaction with the surrounding cloud \citep{gudel18,Tabone2020,LopezVazquez2024}. Sub-mm polarimetry can probe magnetic field geometry in jets and disk winds, although current sensitivity limits such studies to the brightest systems \citep[e.g.,][]{Hull2019}.

ALMA2040 will extend these studies from a few benchmark sources to statistically significant samples spanning stellar mass, evolutionary stage and environment. Sub-au angular resolution, sub-$\mu$Jy continuum sensitivity and broad-band spectral coverage will resolve faint and extended outflow components and enable simultaneous imaging of disks, winds, jets and shocked interaction regions in multiple tracers. Full-Stokes observations will map polarized dust emission and molecular-line polarization, including the Goldreich--Kylafis effect, directly constraining the three-dimensional magnetic field geometry responsible for launching, accelerating and collimating protostellar jets and disk winds \cite{Goldreich1981,Goldreich1982,Hull2019}. Repeated observations will connect episodic accretion to jet response by tracking disturbances from the launch region into the larger-scale flow.

\vspace{0.1cm} \noindent \textbf{AGN: event-horizon-scale jets and particle acceleration.\,\,} Global mm/sub-mm VLBI will be transformative for AGN jet physics (e.g., via Event Horizon Telescope; see companion ALMA2040 white paper \textit{The transformative power of ALMA2040
for mm-VLBI}). With ALMA2040's sensitivity, operations at 345\,GHz and above (ideally to 690\,GHz) become routine, enabling regular imaging of black hole shadows and photon rings in M87* and Sgr~A*, and extending horizon-scale imaging to a broader sample of super-massive black holes (SMBHs) across accretion states and redshifts \citep[e.g.,][]{Ramakrishnan23}. An order-of-magnitude increase in sensitivity would allow detailed polarimetric studies of the jet acceleration and collimation zone within 100 gravitational radii for nearly a dozen additional AGN beyond M87* and Sgr A* \citep{Zhang25}. 

With 5--20 $\mu$as resolution and high sensitivity, candidate SMBH binaries ($10^6-10^9 \, M_{\odot}$) will become directly imageable \cite{Hudson25}, filling a crucial gap between the systems probed by pulsar timing arrays and those expected to be detected by Laser Interferometer Space Antenna (LISA). High-cadence mm polarimetry will trace magnetic field evolution in AGN jets, linking it to gamma-ray flares and neutrino events, and pinpointing the sites of UHE particle acceleration \citep{IceCube_2018}. ALMA2040 thus links horizon-scale magnetic structure to dissipation and particle acceleration by combining sub-THz VLBI with high-cadence polarimetry during multi-messenger alerts.

\vspace{0.1cm} \noindent \textbf{Jetted-TDEs: early mm emission, geometry and outflow demographics.\,\,} TDEs provide a direct view of outflow launching during a short-lived episode of super-Eddington accretion. In on-axis events such as Swift~J1644+57 and AT2022cmc, the radio--mm afterglow is well described as a relativistic outflow interacting with the circumnuclear medium (CNM) \citep{Zauderer2011,Zauderer2013,Berger2012,Metzger2012,Eftekhari2018,Cendes2021}. Millimeter emission can emerge within days and becomes optically thin earlier than cm-band data, making early mm follow-up essential for constraining the initial Lorentz factor, kinetic energy, deceleration time and CNM density profile via the evolving synchrotron peak, as well as particle acceleration physics via the mm spectral index \citep[e.g.,][]{BarniolDuran2013,Eftekhari2018, Rhodes2025}. Polarization measurements add direct constraints on magnetic field ordering and the magneto-ionic CNM; at mm/sub-mm frequencies, contamination from host-galaxy star-formation emission is often reduced relative to cm bands, easing isolation of the nuclear transient component.

A central uncertainty is jet geometry and viewing angle. The long-standing ``increasing energy'' inferred for Swift~J1644+57 from radio equipartition analyses can be explained if the jet was viewed slightly off axis, such that the apparent energy grows as the beaming cone widens during deceleration \citep{Beniamini2023}. This motivates systematic early mm spectral coverage combined with late-time cm--mm monitoring and, for the nearest and brightest events, independent compactness constraints \citep[e.g., VLBI upper limits or marginal size measurements; ][]{Yang2016}. 
ALMA2040’s rapid response and multi-band capability directly address these degeneracies by capturing the earliest mm rise/turnover and anchoring the high-frequency synchrotron spectrum at the epochs most diagnostic of relativistic beaming and magnetization.

Beyond the rare on-axis population, the next frontier is establishing which physical channels power radio emission in the broader TDE population. Synchrotron emission may arise from off-axis relativistic jets, mildly relativistic winds, shocks from stream--stream self-interactions, or shocks involving unbound debris, and these scenarios can produce similar late-time cm-band light curves. Indeed, late-time radio emission on $\sim$500--3000 day timescales is detected in $\sim$40\% of optically selected TDEs \citep[e.g.,][]{Horesh2021,Cendes2022,Cendes2024}. ALMA2040 will enable mm detections and spectra for substantially fainter nuclear transients than currently feasible and will provide statistically controlled measurements of outflow incidence and energetics versus SMBH and host properties. In parallel, molecular gas kinematics in nearby hosts can provide dynamical SMBH masses, turning the dependence of outflow production on SMBH properties and environment into testable empirical relations \citep[e.g.,][]{Nguyen2021,Davis2013}.

\section{ALMA2040: Technical Requirements}
%Summarize the technical requirements here (e.g., angular resolution, spectral resolution, band, bandwidth, sensitivity).
%\subsection{Simulations with GHOST}
%Describe the configurations used to obtain the mock observations from GHOST at \url{https://alma-ghost.jb.man.ac.uk} and present the results.

Achieving the above goals requires major gains in sensitivity, bandwidth, calibration, flexibility, responsiveness and angular resolution. An order-of-magnitude improvement in continuum sensitivity across 35--950 GHz relative to ALMA ($\approx$5--10$\times$ beyond the wideband sensitivity upgrade, WSU) is needed to detect faint jets, capture rapid spectral evolution, reach sub-second time precision and open new regimes of black hole and jet imaging. This requires a 5--7$\times$ larger collecting area (relative to ALMA‑WSU), advanced receiver technology, next‑generation correlators and processing pipelines supporting expanded bandwidths and high time resolution. 

Full‑Stokes capability is essential to recover intrinsic jet orientations, magnetic field structure and Faraday rotation. In the deep Galactic Plane, cm-band polarimetry can be limited by strong Faraday depolarization and $n\pi$ ambiguities, making higher frequency polarimetry necessary to measure rotation measures and magnetic field evolution robustly. Such measurements are sensitivity hungry and require high signal-to-noise detections across multiple bands. Multi‑band receivers and flexible sub‑array operation are therefore needed for simultaneous spectral and polarimetric coverage during high-cadence campaigns, efficient array use and agile ToO response. Sub-arrays can also provide a high-sensitivity ground anchor for future mm-space-VLBI missions while maintaining a conventional science schedule. A fully automated, minute-to-hour-scale response system is crucial for GRB and other multi‑messenger triggers. 

Stable long‑baseline performance and accurate astrometry at the diffraction limit are vital for jet‑motion studies, detection of orbital wobble in SMBH binaries, and compact object tracking in Galactic systems, requiring $\mu$as‑level phase stability. For rapidly moving ejecta (e.g., XRB jets), standard Earth-rotation synthesis imaging can break down as components traverse many synthesized beams during a typical track; ALMA2040 should therefore support motion-aware calibration/imaging (or visibility-domain modeling) within VLBI and long-baseline pipelines. Baselines of $\approx$50 km and full compatibility with global mm/sub‑mm VLBI networks operating at $\geq$345\,GHz (ideally 690\,GHz) are essential for next-generation black hole imaging (see more detailed information in the companion ALMA2040 white paper \textit{The transformative power of ALMA2040 for mm-VLBI}).

A core goal of ALMA2040 is to move transient and variable jet studies from ``case studies'' to statistically controlled follow-up samples by making triggered observations fast and sensitive enough that non-detections become constraining and fainter/off-axis populations become routinely measurable. This requires four complementary observing modes:

\vspace{0.1cm} \noindent \textbf{\textbullet\, Rapid-response mode} with automated triggers and on-source times of $\lesssim$10--30 minutes for GRBs and fast transients, and $\lesssim$hours for neutrino/gamma-ray alerts. For GRBs, the reverse shock and afterglow onset at mm/sub-mm frequencies often evolve on minute-to-hour timescales, implying that response time is as important as raw sensitivity.

\vspace{0.1cm} \noindent \textbf{\textbullet\, High-cadence monitoring mode} for Galactic systems (XRBs, microquasars) during outburst, enabling second-scale integration times and polarimetry to track evolving jet breaks, magnetic field rotations and shock propagation.

\vspace{0.1cm} \noindent \textbf{\textbullet\, Subarray mode} will allow agile deployment to support extreme-resolution VLBI followups of GRB candidates, high energy neutrino sources and rapidly evolving merger afterglows. Subarrays can also be used to record across a tremendous frequency range using different multi-frequency configurations. On-the-fly scheduling optimization routine would also allow to delegate targeting of bright point sources only to a subset of telescopes.

\vspace{0.1cm} \noindent \textbf{\textbullet\, VLBI mode} is necessary to achieve the extreme resolution necessary to resolve black hole shadows and the jet launching region in AGN systems. Participation of ALMA2040 in simultaneous multi-frequency VLBI observations at 86/230/345GHz will allow for Faraday rotation and spectral index mapping on $\mu as$ scales, giving us direct probes of magnetic field properties immediately surrounding the black hole. Sub-THz VLBI will probe the photon ring scale in M87* and Sgr A*, and an upgraded phasing system to support longer coherence times at 690GHz will also improve coherence and sensitivity at lower frequencies.  

%
%\section{Synergies with current and future facilities}
%This section is optional. 
\section{Synergies with current and future facilities}

\noindent \textbf{Synergies with cm--mm radio facilities (SKAO, ngVLA, DSA)\,\,} The combination of ALMA2040 with next-generation cm/mm radio facilities will provide a multi-scale view of the link between relativistic jet launching and galaxy-scale feedback. ALMA2040 will probe the innermost jet, where the flow is launched, accelerated and collimated, with a $\gtrsim10\times$ gain in continuum sensitivity, $35$--$950$~GHz frequency coverage, high-fidelity polarimetry and $\sim50$~km baselines. At lower frequencies, SKAO and ngVLA will trace the subsequent propagation and interaction of these jets with the ISM through sensitive continuum imaging, spectral ageing, and (via VLBI) spatially resolved H{\sc i} absorption \citep[e.g.,][]{Oosterloo2023, Morganti2004, Schulz2018}. Combined with ALMA2040 CO and dense gas observations, these data will determine whether jets heat and disrupt atomic gas, compress it into molecular gas, or drive turbulence that suppresses star formation \citep[e.g.,][]{Pandey-Pommier01.2026.SKA,Pandey-Pommier03.2026.SKA}.

ngVLA coverage of $\sim$1.2--116\,GHz with large instantaneous bandwidth and long baselines traces optically thick synchrotron emission and longer-lived ejecta at lower frequencies. Joint ngVLA and ALMA2040 campaigns will measure evolving jet spectra and time lags across cm to sub-mm bands to constrain jet speeds, opening angles and particle acceleration physics in XRB outbursts and TDEs. For multi-messenger and fast-transient science, wide-field radio mapping of large localization regions to $\mu$Jy depths---including with facilities such as the Deep Synoptic Array (DSA)---enables prompt source localization and characterization that can be followed immediately by ALMA2040’s high-frequency photometry and polarimetry. 

\vspace{0.1cm} \noindent \textbf{Synergies with mm/sub-mm facilities (wSMA, AtLAST and CMB surveys)\,\,} The ongoing Submillimeter Array wideband upgrade (wSMA) will act as a pathfinder for wideband mm observing modes, time-domain operations and rapid-response ToOs, while ALMA2040 extends these capabilities with substantially higher sensitivity, longer baselines and full-Stokes polarimetry \citep[e.g.,][]{wSMA2024}. Wide-field CMB surveys (e.g., South Pole Telescope, Simons Observatory, and successors in the 2040s, as well as future satellite missions like CMB Probe+) can serve as all-sky mm monitors for bright transients by covering large fractions of the sky per day, albeit with modest localization (typically $\sim$10$''$). Large single-dish facilities, such as the Atacama Large Aperture Submillimeter Telescope (AtLAST) can map large error regions from multi-messenger alerts, while ALMA2040 follow-up provides precise positions for secure multi-wavelength association together with multi-band spectra and polarimetry \citep[e.g.,][]{Tandoi2024, Biermann2025, Tandoi2026, 2022ApJ...935...16E}.

\vspace{0.1cm} \noindent \textbf{Synergies with optical/UV/X-ray observatories\,\,} Wide-field optical surveys (e.g., Argus Array) and UV transient discovery (e.g., ULTRASAT) will provide triggers for ALMA2040 ToOs, while next-generation X-ray missions (e.g., NewAthena) will provide prompt characterization of accretion and high-energy emission. High-speed optical photometry with ELT-class ($\sim$30\,m) telescopes on sub-second to minute timescales will complement mm-band timing (Figure~\ref{fig:xrb}), enabling measurements of inter-band lags and correlated variability that trace the propagation of fluctuations through the inflow--outflow system \citep[e.g.,][]{Gandhi2010}.

\vspace{0.1cm} \noindent \textbf{Synergies with multi-messenger observatories\,\,} Third-generation GW detectors (Einstein Telescope, Cosmic Explorer) will provide large samples of compact-object mergers out to cosmological distances, while LISA will probe SMBH binaries and extreme-mass-ratio inspirals. For the subset of mergers that launch relativistic outflows, mm/sub-mm afterglows will provide early-time calorimetry and strong constraints on jet angular structure and viewing angle, as demonstrated by GW170817 \citep[e.g.,][]{Mooley2018,Alexander2018,Margutti2018}. Population studies predict that most GW-triggered afterglows will be observed off-axis and will often be faint, motivating sub-$\mu$Jy sensitivity and rapid ToO response to build unbiased samples across distance and inclination \citep[e.g.,][]{Hotokezaka2015}. Neutrino observatories (e.g., IceCube-Gen2 and KM3NeT), together with next-generation UHE gamma-ray facilities (CTAO and SWGO), will deliver real-time alerts for candidate hadronic accelerators. Rapid mm/sub-mm spectral and polarization follow-up can localize dissipation sites, constrain magnetization through rotation measures and depolarization, and test whether the same episodes that produce high-energy neutrinos also drive mm-band shocks or reconnection events \citep[e.g.,][]{IceCube_2018, Hovatta2012}. Achieving this in unbiased samples will require minute-to-hour ToO response and substantially improved sensitivity and polarimetric capabilities in the mm-band.

\section*{Acknowledgements}
We sincerely thank James Miller-Jones (Curtin University) and Michael McCollough (Harvard-Smithsonian Center for Astrophysics) for reviewing this paper. This work was developed as part of the broader ALMA2040 community effort. We acknowledge the Lorentz Center and NOVA for supporting the community workshops and coordination activities that helped initiate and shape the ALMA2040 White Paper process.

\newpage
\bibliographystyle{apj_mod}
\noindent {\bibliography{references}}

@ARTICLE{mal14,
       author = {{Malzac}, Julien},
        title = "{The spectral energy distribution of compact jets powered by internal shocks}",
      journal = {MNRAS},
         year = 2014,
        month = sep,
       volume = {443},
       number = {1},
        pages = {299-317},
          doi = {10.1093/mnras/stu1144},
archivePrefix = {arXiv},
       eprint = {1406.2208},
 primaryClass = {astro-ph.HE},
       adsurl = {https://ui.adsabs.harvard.edu/abs/2014MNRAS.443..299M}
}

@ARTICLE{gan11,
       author = {{Gandhi}, P. and {Blain}, A.~W. and {Russell}, D.~M. and {Casella}, P. and {Malzac}, J. and {Corbel}, S. and {D'Avanzo}, P. and {Lewis}, F.~W. and {Markoff}, S. and {Cadolle Bel}, M. and {Goldoni}, P. and {Wachter}, S. and {Khangulyan}, D. and {Mainzer}, A.},
        title = "{A Variable Mid-infrared Synchrotron Break Associated with the Compact Jet in GX 339-4}",
      journal = {ApJL},
         year = 2011,
        month = oct,
       volume = {740},
       number = {1},
          eid = {L13},
        pages = {L13},
          doi = {10.1088/2041-8205/740/1/L13},
archivePrefix = {arXiv},
       eprint = {1109.4143},
 primaryClass = {astro-ph.HE},
       adsurl = {https://ui.adsabs.harvard.edu/abs/2011ApJ...740L..13G}
}

@ARTICLE{Morganti2004,
       author = {{Morganti}, R. and {Oosterloo}, T.~A. and {Tadhunter}, C.~N. and {Vermeulen}, R. and {Pihlstr{\"o}m}, Y.~M. and {van Moorsel}, G. and {Wills}, K.~A.},
        title = "{The unfriendly ISM in the radio  galaxy 4C 12.50 (PKS 1345+12)}",
      journal = {\aap},
         year = 2004,
        month = sep,
       volume = {424},
        pages = {119-124},
          doi = {10.1051/0004-6361:20041064},
archivePrefix = {arXiv},
       eprint = {astro-ph/0406012},
 primaryClass = {astro-ph},
       adsurl = {https://ui.adsabs.harvard.edu/abs/2004A&A...424..119M}
}

@ARTICLE{2022ApJ...935...16E,
       author = {{Eftekhari}, T. and {Berger}, E. and {Metzger}, B.~D. and {Laskar}, T. and {Villar}, V.~A. and {Alexander}, K.~D. and {Holder}, G.~P. and {Vieira}, J.~D. and {Whitehorn}, N. and {Williams}, P.~K.~G.},
        title = "{Extragalactic Millimeter Transients in the Era of Next-generation CMB Surveys}",
      journal = {\apj},
         year = 2022,
        month = aug,
       volume = {935},
       number = {1},
          eid = {16},
        pages = {16},
          doi = {10.3847/1538-4357/ac7ce8},
archivePrefix = {arXiv},
       eprint = {2110.05494},
 primaryClass = {astro-ph.HE},
       adsurl = {https://ui.adsabs.harvard.edu/abs/2022ApJ...935...16E}
}

@ARTICLE{Tandoi2024,
       author = {{Tandoi}, C. and {Guns}, S. and {Foster}, A. and {Ade}, P.~A.~R. and {Anderson}, A.~J. and {Ansarinejad}, B. and {Archipley}, M. and {Balkenhol}, L. and {Benabed}, K. and {Bender}, A.~N. and {Benson}, B.~A. and {Bianchini}, F. and {Bleem}, L.~E. and {Bouchet}, F.~R. and {Bryant}, L. and {Camphuis}, E. and {Carlstrom}, J.~E. and {Cecil}, T.~W. and {Chang}, C.~L. and {Chaubal}, P. and {Chichura}, P.~M. and {Chou}, T.-L. and {Coerver}, A. and {Crawford}, T.~M. and {Cukierman}, A. and {Daley}, C. and {de Haan}, T. and {Dibert}, K.~R. and {Dobbs}, M.~A. and {Doussot}, A. and {Dutcher}, D. and {Everett}, W. and {Feng}, C. and {Ferguson}, K.~R. and {Fichman}, K. and {Galli}, S. and {Gambrel}, A.~E. and {Gardner}, R.~W. and {Ge}, F. and {Goeckner-Wald}, N. and {Gualtieri}, R. and {Guidi}, F. and {Halverson}, N.~W. and {Hivon}, E. and {Holder}, G.~P. and {Holzapfel}, W.~L. and {Hood}, J.~C. and {Huang}, N. and {K{\'e}ruzor{\'e}}, F. and {Knox}, L. and {Korman}, M. and {Kornoelje}, K. and {Kuo}, C.-L. and {Lee}, A.~T. and {Levy}, K. and {Lowitz}, A.~E. and {Lu}, C. and {Maniyar}, A. and {Menanteau}, F. and {Millea}, M. and {Montgomery}, J. and {Moon}, Y. and {Nakato}, Y. and {Natoli}, T. and {Noble}, G.~I. and {Novosad}, V. and {Omori}, Y. and {Padin}, S. and {Pan}, Z. and {Paschos}, P. and {Phadke}, K.~A. and {Prabhu}, K. and {Qu}, Z. and {Quan}, W. and {Rahimi}, M. and {Rahlin}, A. and {Reichardt}, C.~L. and {Reuter}, C. and {Rouble}, M. and {Ruhl}, J.~E. and {Schiappucci}, E. and {Smecher}, G. and {Sobrin}, J.~A. and {Stark}, A.~A. and {Stephen}, J. and {Suzuki}, A. and {Thompson}, K.~L. and {Thorne}, B. and {Trendafilova}, C. and {Tucker}, C. and {Umilta}, C. and {Vieira}, J.~D. and {Wan}, Y. and {Wang}, G. and {Whitehorn}, N. and {Wu}, W.~L.~K. and {Yefremenko}, V. and {Young}, M.~R. and {Zebrowski}, J.~A.},
        title = "{Flaring Stars in a Nontargeted Millimeter-wave Survey with SPT-3G}",
      journal = {\apj},
         year = 2024,
        month = sep,
       volume = {972},
       number = {1},
          eid = {6},
        pages = {6},
          doi = {10.3847/1538-4357/ad58db},
archivePrefix = {arXiv},
       eprint = {2401.13525},
 primaryClass = {astro-ph.SR},
       adsurl = {https://ui.adsabs.harvard.edu/abs/2024ApJ...972....6T}
}

@ARTICLE{Tandoi2026,
       author = {{Tandoi}, C. and {Foster}, A. and {Maccarone}, T.~J. and {Anderson}, A.~J. and {Ansarinejad}, B. and {Archipley}, M. and {Balkenhol}, L. and {Barron}, D.~R. and {Benabed}, K. and {Bender}, A.~N. and {Benson}, B.~A. and {Bianchini}, F. and {Bleem}, L.~E. and {Bocquet}, S. and {Bouchet}, F.~R. and {Camphuis}, E. and {Campitiello}, M.~G. and {Carlstrom}, J.~E. and {Carron}, J. and {Chang}, C.~L. and {Chichura}, P.~M. and {Chokshi}, A. and {Chou}, T.-L. and {Coerver}, A. and {Crawford}, T.~M. and {Daley}, C. and {de Haan}, T. and {Dibert}, K.~R. and {Dobbs}, M.~A. and {Doohan}, M. and {Dutcher}, D. and {Feng}, C. and {Ferguson}, K.~R. and {Ferree}, N.~C. and {Fichman}, K. and {Galli}, S. and {Gambrel}, A.~E. and {Gao}, A.~K. and {Ge}, F. and {Guidi}, F. and {Guns}, S. and {Halverson}, N.~W. and {Hivon}, E. and {Holder}, G.~P. and {Holzapfel}, W.~L. and {Hood}, J.~C. and {Hryciuk}, A. and {Huang}, N. and {Jhaveri}, T. and {Keruzore}, F. and {Khalife}, A.~R. and {Knox}, L. and {Kornoelje}, K. and {Kuo}, C.-L. and {Levy}, K. and {Li}, Y. and {Lowitz}, A.~E. and {Lu}, C. and {Lynch}, G.~P. and {Maniyar}, A.~S. and {Martsen}, E.~S. and {Menanteau}, F. and {Millea}, M. and {Montgomery}, J. and {Nakato}, Y. and {Natoli}, T. and {Ouellette}, A. and {Pan}, Z. and {Paschos}, P. and {Phadke}, K.~A. and {Pollak}, A.~W. and {Prabhu}, K. and {Quan}, W. and {Rahimi}, M. and {Rahlin}, A. and {Reichardt}, C.~L. and {Rouble}, M. and {Ruhl}, J.~E. and {Silva Oliveira}, A.~C. and {Simpson}, A. and {Sobrin}, J.~A. and {Stark}, A.~A. and {Stephen}, J. and {Trendafilova}, C. and {Vieira}, J.~D. and {Vieregg}, A.~G. and {Vitrier}, A. and {Wan}, Y. and {Whitehorn}, N. and {Wu}, W.~L.~K. and {Young}, M.~R. and {Zebrowski}, J.~A.},
        title = "{Millimeter-wave Detections of Symbiotic Stars in SPT and ACT Data}",
      journal = {arXiv e-prints},
         year = 2026,
        month = may,
          eid = {arXiv:2605.01022},
        pages = {arXiv:2605.01022},
          doi = {10.48550/arXiv.2605.01022},
archivePrefix = {arXiv},
       eprint = {2605.01022},
 primaryClass = {astro-ph.SR},
       adsurl = {https://ui.adsabs.harvard.edu/abs/2026arXiv260501022T}
}

@ARTICLE{Biermann2025,
       author = {{Biermann}, Emily and {Li}, Yaqiong and {Naess}, Sigurd and {Choi}, Steve K. and {Clark}, Susan E. and {Devlin}, Mark and {Dunkley}, Jo and {Gallardo}, P.~A. and {Guan}, Yilun and {Foster}, Allen and {Hasselfield}, Matthew and {Herv{\'\i}as-Caimapo}, Carlos and {Hilton}, Matt and {Hincks}, Adam D. and {Ho}, Anna Y.~Q. and {Hood}, John C. and {Huffenberger}, Kevin M. and {Kosowsky}, Arthur and {Niemack}, Michael D. and {Orlowski-Scherer}, John and {Page}, Lyman and {Partridge}, Bruce and {Salatino}, Maria and {Sif{\'o}n}, Crist{\'o}bal and {Staggs}, Suzanne T. and {Vargas}, Cristian and {Wollack}, Edward J.},
        title = "{The Atacama Cosmology Telescope: Systematic Transient Search of Single Observation Maps}",
      journal = {\apj},
         year = 2025,
        month = jun,
       volume = {986},
       number = {1},
          eid = {7},
        pages = {7},
          doi = {10.3847/1538-4357/adce70},
archivePrefix = {arXiv},
       eprint = {2409.08429},
 primaryClass = {astro-ph.HE},
       adsurl = {https://ui.adsabs.harvard.edu/abs/2025ApJ...986....7B}
}

@INCOLLECTION{Pandey-Pommier01.2026.SKA,
       author = {{Pandey-Pommier}, Mamta and {Sethuram}, Ramya and {Roy}, Subhashis and {Muthumariappan}, Chinnathambi and {Roy}, Nirupam and {Marcowith}, Alexandre and {Konar}, Chiranjib},
        title = "{Tracing Gas Outflows in Molecular Hydrogen Emission-rich Galaxies with the SKA}",
    booktitle = {Advancing Astrophysics with the SKA II},
         year = 2026,
       editor = {},
          eid = {88},
        pages = {88},
       adsurl = {https://ui.adsabs.harvard.edu/abs/2026aask.bookE..88P}
}

@INCOLLECTION{Pandey-Pommier03.2026.SKA,
       author = {{Pandey-Pommier}, Mamta},
        title = "{Uncovering Neutral Hydrogen Clouds in Radio Galaxies}",
    booktitle = {Advancing Astrophysics with the SKA I},
         year = 2026,
       editor = {},
          eid = {66},
        pages = {66},
          doi = {10.48550/arXiv.2606.31773},
       adsurl = {https://ui.adsabs.harvard.edu/abs/2026aask.bookE..66P}
}

@ARTICLE{Oosterloo2023,
       author = {{Oosterloo}, Tom and {Morganti}, Raffaella and {Murthy}, Suma},
        title = "{Closing the feedback-feeding loop of the radio galaxy 3C 84}",
      journal = {Nature Astronomy},
         year = 2024,
        month = feb,
       volume = {8},
        pages = {256-262},
          doi = {10.1038/s41550-023-02138-y},
archivePrefix = {arXiv},
       eprint = {2312.00917},
 primaryClass = {astro-ph.GA},
       adsurl = {https://ui.adsabs.harvard.edu/abs/2024NatAs...8..256O}
}

@ARTICLE{Schulz2018,
       author = {{Schulz}, R. and {Morganti}, R. and {Nyland}, K. and {Paragi}, Z. and {Mahony}, E.~K. and {Oosterloo}, T.},
        title = "{Mapping the neutral atomic hydrogen gas outflow in the restarted radio galaxy 3C 236}",
      journal = {\aap},
         year = 2018,
        month = sep,
       volume = {617},
          eid = {A38},
        pages = {A38},
          doi = {10.1051/0004-6361/201833108},
archivePrefix = {arXiv},
       eprint = {1806.06653},
 primaryClass = {astro-ph.GA},
       adsurl = {https://ui.adsabs.harvard.edu/abs/2018A&A...617A..38S}
}

@ARTICLE{tetarenko18,
       author = {{Tetarenko, A.~J.} and {Freeman}, P. and {Rosolowsky}, E.~W. and {Miller-Jones}, J.~C.~A. and {Sivakoff}, G.~R.},
        title = "{Mapping jet-ISM interactions in X-ray binaries with ALMA: a GRS 1915+105 case study}",
      journal = {\mnras},
         year = 2018,
        month = mar,
       volume = {475},
       number = {1},
        pages = {448-468},
          doi = {10.1093/mnras/stx3151},
archivePrefix = {arXiv},
       eprint = {1712.00432},
 primaryClass = {astro-ph.HE},
       adsurl = {https://ui.adsabs.harvard.edu/abs/2018MNRAS.475..448T}
}

@ARTICLE{Nguyen2021,
       author = {{Nguyen}, Dieu D. and {Izumi}, Takuma and {Thater}, Sabine and {Imanishi}, Masatoshi and {Kawamuro}, Taiki and {Baba}, Shunsuke and {Nakano}, Suzuka and {Turner}, Jean L. and {Kohno}, Kotaro and {Matsushita}, Satoki and {Mart{\'\i}n}, Sergio and {Meier}, David S. and {Nguyen}, Phuong M. and {Nguyen}, Lam T.},
        title = "{Black hole mass measurement using ALMA observations of [CI] and CO emissions in the Seyfert 1 galaxy NGC 7469}",
      journal = {\mnras},
         year = 2021,
        month = jul,
       volume = {504},
       number = {3},
        pages = {4123-4142},
          doi = {10.1093/mnras/stab1002},
archivePrefix = {arXiv},
       eprint = {2104.03539},
 primaryClass = {astro-ph.GA},
       adsurl = {https://ui.adsabs.harvard.edu/abs/2021MNRAS.504.4123N}
}

@ARTICLE{Davis2013,
       author = {{Davis}, Timothy A. and {Bureau}, Martin and {Cappellari}, Michele and {Sarzi}, Marc and {Blitz}, Leo},
        title = "{A black-hole mass measurement from molecular gas kinematics in NGC4526}",
      journal = {\nat},
         year = 2013,
        month = feb,
       volume = {494},
       number = {7437},
        pages = {328-330},
          doi = {10.1038/nature11819},
archivePrefix = {arXiv},
       eprint = {1301.7184},
 primaryClass = {astro-ph.CO},
       adsurl = {https://ui.adsabs.harvard.edu/abs/2013Natur.494..328D}
}

@ARTICLE{tetarenko21,
       author = {{Tetarenko, A.~J.} and {Casella}, P. and {Miller-Jones}, J.~C.~A. and {Sivakoff}, G.~R. and {Paice}, J.~A. and {Vincentelli}, F.~M. and {Maccarone}, T.~J. and {Gandhi}, P. and {Dhillon}, V.~S. and {Marsh}, T.~R. and {Russell}, T.~D. and {Uttley}, P.},
        title = "{Measuring fundamental jet properties with multiwavelength fast timing of the black hole X-ray binary MAXI J1820+070}",
      journal = {\mnras},
         year = 2021,
        month = jul,
       volume = {504},
       number = {3},
        pages = {3862-3883},
          doi = {10.1093/mnras/stab820},
archivePrefix = {arXiv},
       eprint = {2103.09318},
 primaryClass = {astro-ph.HE},
       adsurl = {https://ui.adsabs.harvard.edu/abs/2021MNRAS.504.3862T}
}

@ARTICLE{zdziarski22,
       author = {{Zdziarski}, Andrzej A. and {Tetarenko}, Alexandra J. and {Sikora}, Marek},
        title = "{Jet Parameters in the Black Hole X-Ray Binary MAXI J1820+070}",
      journal = {\apj},
         year = 2022,
        month = feb,
       volume = {925},
       number = {2},
          eid = {189},
        pages = {189},
          doi = {10.3847/1538-4357/ac38a9},
archivePrefix = {arXiv},
       eprint = {2108.10929},
 primaryClass = {astro-ph.HE},
       adsurl = {https://ui.adsabs.harvard.edu/abs/2022ApJ...925..189Z}
}

@ARTICLE{vandeneijnden22,
       author = {{van den Eijnden}, J. and {Degenaar}, N. and {Russell}, T.~D. and {Miller-Jones}, J.~C.~A. and {Rouco Escorial}, A. and {Wijnands}, R. and {Sivakoff}, G.~R. and {Hern{\'a}ndez Santisteban}, J.~V.},
        title = "{Radio monitoring of transient Be/X-ray binaries and the inflow-outflow coupling of strongly magnetized accreting neutron stars}",
      journal = {\mnras},
         year = 2022,
        month = nov,
       volume = {516},
       number = {4},
        pages = {4844-4861},
          doi = {10.1093/mnras/stac2518},
archivePrefix = {arXiv},
       eprint = {2208.14903},
 primaryClass = {astro-ph.HE},
       adsurl = {https://ui.adsabs.harvard.edu/abs/2022MNRAS.516.4844V}
}

@ARTICLE{baglio23,
       author = {{Baglio}, M.~C. and {Coti Zelati}, F. and {Campana}, S. and {Busquet}, G. and {D'Avanzo}, P. and {Giarratana}, S. and {Giroletti}, M. and {Ambrosino}, F. and {Crespi}, S. and {Miraval Zanon}, A. and {Hou}, X. and {Li}, D. and {Li}, J. and {Wang}, P. and {Russell}, D.~M. and {Torres}, D.~F. and {Alabarta}, K. and {Casella}, P. and {Covino}, S. and {Bramich}, D.~M. and {de Martino}, D. and {M{\'e}ndez}, M. and {Motta}, S.~E. and {Papitto}, A. and {Saikia}, P. and {Vincentelli}, F.},
        title = "{Matter ejections behind the highs and lows of the transitional millisecond pulsar PSR J1023+0038}",
      journal = {\aap},
         year = 2023,
        month = sep,
       volume = {677},
          eid = {A30},
        pages = {A30},
          doi = {10.1051/0004-6361/202346418},
archivePrefix = {arXiv},
       eprint = {2305.14509},
 primaryClass = {astro-ph.HE},
       adsurl = {https://ui.adsabs.harvard.edu/abs/2023A&A...677A..30B}
}

@ARTICLE{koljonen25,
       author = {{Koljonen}, K.~I.~I. and {Linares}, M. and {Miller-Jones}, J.~C.~A.},
        title = "{Jet break revealed in the transitional millisecond pulsar candidate 4FGL J0427.8-6704}",
      journal = {\mnras},
         year = 2025,
        month = may,
       volume = {539},
       number = {1},
        pages = {95-108},
          doi = {10.1093/mnras/staf483},
archivePrefix = {arXiv},
       eprint = {2503.20902},
 primaryClass = {astro-ph.HE},
       adsurl = {https://ui.adsabs.harvard.edu/abs/2025MNRAS.539...95K}
}

@ARTICLE{russell24,
       author = {{Russell}, Thomas D. and {Degenaar}, Nathalie and {van den Eijnden}, Jakob and {Maccarone}, Thomas and {Tetarenko}, Alexandra J. and {S{\'a}nchez-Fern{\'a}ndez}, Celia and {Miller-Jones}, James C.~A. and {Kuulkers}, Erik and {Del Santo}, Melania},
        title = "{Thermonuclear explosions on neutron stars reveal the speed of their jets}",
      journal = {\nat},
         year = 2024,
        month = mar,
       volume = {627},
       number = {8005},
        pages = {763-766},
          doi = {10.1038/s41586-024-07133-5},
archivePrefix = {arXiv},
       eprint = {2403.18135},
 primaryClass = {astro-ph.HE},
       adsurl = {https://ui.adsabs.harvard.edu/abs/2024Natur.627..763R}
}

@ARTICLE{lhaaso24,
       author = {{LHAASO Collaboration}},
        title = "{Ultrahigh-Energy Gamma-ray Emission Associated with Black Hole-Jet Systems}",
      journal = {arXiv e-prints},
         year = 2024,
        month = oct,
          eid = {arXiv:2410.08988},
        pages = {arXiv:2410.08988},
          doi = {10.48550/arXiv.2410.08988},
archivePrefix = {arXiv},
       eprint = {2410.08988},
 primaryClass = {astro-ph.HE},
       adsurl = {https://ui.adsabs.harvard.edu/abs/2024arXiv241008988L}
}

@ARTICLE{koljonen23,
       author = {{Koljonen}, Karri I.~I. and {Satalecka}, Konstancja and {Lindfors}, Elina J. and {Liodakis}, Ioannis},
        title = "{Microquasar Cyg X-3 - a unique jet-wind neutrino factory?}",
      journal = {\mnras},
         year = 2023,
        month = sep,
       volume = {524},
       number = {1},
        pages = {L89-L93},
          doi = {10.1093/mnrasl/slad081},
archivePrefix = {arXiv},
       eprint = {2306.11804},
 primaryClass = {astro-ph.HE},
       adsurl = {https://ui.adsabs.harvard.edu/abs/2023MNRAS.524L..89K}
}

@ARTICLE{zhang05,
       author = {{Zhang}, Bing and {Kobayashi}, Shiho},
        title = "{Gamma-Ray Burst Early Afterglows: Reverse Shock Emission from an Arbitrarily Magnetized Ejecta}",
      journal = {\apj},
         year = 2005,
        month = jul,
       volume = {628},
       number = {1},
        pages = {315-334},
          doi = {10.1086/429787},
archivePrefix = {arXiv},
       eprint = {astro-ph/0404140},
 primaryClass = {astro-ph},
       adsurl = {https://ui.adsabs.harvard.edu/abs/2005ApJ...628..315Z}
}

@ARTICLE{granot02,
       author = {{Granot}, Jonathan and {Sari}, Re'em},
        title = "{The Shape of Spectral Breaks in Gamma-Ray Burst Afterglows}",
      journal = {\apj},
         year = 2002,
        month = apr,
       volume = {568},
       number = {2},
        pages = {820-829},
          doi = {10.1086/338966},
archivePrefix = {arXiv},
       eprint = {astro-ph/0108027},
 primaryClass = {astro-ph},
       adsurl = {https://ui.adsabs.harvard.edu/abs/2002ApJ...568..820G}
}

@ARTICLE{gao13,
       author = {{Gao}, He and {Ding}, Xuan and {Wu}, Xue-Feng and {Zhang}, Bing and {Dai}, Zi-Gao},
        title = "{Bright Broadband Afterglows of Gravitational Wave Bursts from Mergers of Binary Neutron Stars}",
      journal = {\apj},
         year = 2013,
        month = jul,
       volume = {771},
       number = {2},
          eid = {86},
        pages = {86},
          doi = {10.1088/0004-637X/771/2/86},
archivePrefix = {arXiv},
       eprint = {1301.0439},
 primaryClass = {astro-ph.HE},
       adsurl = {https://ui.adsabs.harvard.edu/abs/2013ApJ...771...86G}
}

@ARTICLE{laskar19,
       author = {{Laskar}, Tanmoy and {Alexander}, Kate D. and {Gill}, Ramandeep and {Granot}, Jonathan and {Berger}, Edo and {Mundell}, C.~G. and {Barniol Duran}, Rodolfo and {Bolmer}, J. and {Duffell}, Paul and {van Eerten}, Hendrik and {Fong}, Wen-fai and {Kobayashi}, Shiho and {Margutti}, Raffaella and {Schady}, Patricia},
        title = "{ALMA Detection of a Linearly Polarized Reverse Shock in GRB 190114C}",
      journal = {\apjl},
         year = 2019,
        month = jun,
       volume = {878},
       number = {1},
          eid = {L26},
        pages = {L26},
          doi = {10.3847/2041-8213/ab2247},
archivePrefix = {arXiv},
       eprint = {1904.07261},
 primaryClass = {astro-ph.HE},
       adsurl = {https://ui.adsabs.harvard.edu/abs/2019ApJ...878L..26L}
}

@ARTICLE{Ramakrishnan23,
       author = {{Ramakrishnan}, Venkatessh and {Nagar}, Neil and {Arratia}, Vicente and {Hern{\'a}ndez-Y{\'e}venes}, Joaqu{\'\i}n and {Pesce}, Dominic W. and {Nair}, Dhanya G. and {Bandyopadhyay}, Bidisha and {Medina-Porcile}, Catalina and {Krichbaum}, Thomas P. and {Doeleman}, Sheperd and {Ricarte}, Angelo and {Fish}, Vincent L. and {Blackburn}, Lindy and {Falcke}, Heino and {Bower}, Geoffrey and {Natarajan}, Priyamvada},
        title = "{Event Horizon and Environs (ETHER): A Curated Database for EHT and ngEHT Targets and Science}",
      journal = {Galaxies},
         year = 2023,
        month = jan,
       volume = {11},
       number = {1},
          eid = {15},
        pages = {15},
          doi = {10.3390/galaxies11010015},
       adsurl = {https://ui.adsabs.harvard.edu/abs/2023Galax..11...15R}
}

@ARTICLE{Zhang25,
       author = {{Zhang}, Xinyue Alice and {Ricarte}, Angelo and {Pesce}, Dominic W. and {Johnson}, Michael D. and {Nagar}, Neil and {Narayan}, Ramesh and {Ramakrishnan}, Venkatessh and {Doeleman}, Sheperd and {Palumbo}, Daniel C.~M.},
        title = "{Accessing a New Population of Supermassive Black Holes with Extensions to the Event Horizon Telescope}",
      journal = {\apj},
         year = 2025,
        month = may,
       volume = {985},
       number = {1},
          eid = {41},
        pages = {41},
          doi = {10.3847/1538-4357/adbd45},
archivePrefix = {arXiv},
       eprint = {2406.17754},
 primaryClass = {astro-ph.HE},
       adsurl = {https://ui.adsabs.harvard.edu/abs/2025ApJ...985...41Z}
}

@ARTICLE{Hudson25,
       author = {{Hudson}, B. and {Gurvits}, L.~I. and {Mooij}, E. and {Ricarte}, A. and {Palumbo}, D.},
        title = "{Towards direct imaging and orbital parameter estimation of supermassive black hole binaries with spaceborne VLBI}",
      journal = {arXiv e-prints},
         year = 2025,
        month = nov,
          eid = {arXiv:2511.07996},
        pages = {arXiv:2511.07996},
          doi = {10.48550/arXiv.2511.07996},
archivePrefix = {arXiv},
       eprint = {2511.07996},
 primaryClass = {astro-ph.IM},
       adsurl = {https://ui.adsabs.harvard.edu/abs/2025arXiv251107996H}
}

@ARTICLE{gudel18,
       author = {{G{\"u}del}, M. and {Eibensteiner}, C. and {Dionatos}, O. and {Audard}, M. and {Forbrich}, J. and {Kraus}, S. and {Rab}, Ch. and {Schneider}, Ch. and {Skinner}, S. and {Vorobyov}, E.},
        title = "{ALMA detects a radial disk wind in DG Tauri}",
      journal = {\aap},
         year = 2018,
        month = dec,
       volume = {620},
          eid = {L1},
        pages = {L1},
          doi = {10.1051/0004-6361/201834271},
archivePrefix = {arXiv},
       eprint = {1810.12251},
 primaryClass = {astro-ph.SR},
       adsurl = {https://ui.adsabs.harvard.edu/abs/2018A&A...620L...1G}
}

@ARTICLE{neronov25,
       author = {{Neronov}, A. and {Oikonomou}, F. and {Semikoz}, D.},
        title = "{Multimessenger signature of cosmic rays from the microquasar V4641 Sgr propagating along a Galactic magnetic field line}",
      journal = {\prd},
         year = 2025,
        month = may,
       volume = {111},
       number = {10},
          eid = {103025},
        pages = {103025},
          doi = {10.1103/PhysRevD.111.103025},
archivePrefix = {arXiv},
       eprint = {2410.17608},
 primaryClass = {astro-ph.HE},
       adsurl = {https://ui.adsabs.harvard.edu/abs/2025PhRvD.111j3025N}
}

@ARTICLE{bright2023,
       author = {{Bright J. S. \& Rhodes L.}  and {Farah}, Wael and {Fender}, Rob and {van der Horst}, Alexander J. and {Leung}, James K. and {Williams}, David R.~A. and {Anderson}, Gemma E. and {Atri}, Pikky and {DeBoer}, David R. and {Giarratana}, Stefano and {Green}, David A. and {Heywood}, Ian and {Lenc}, Emil and {Murphy}, Tara and {Pollak}, Alexander W. and {Premnath}, Pranav H. and {Scott}, Paul F. and {Sheikh}, Sofia Z. and {Siemion}, Andrew and {Titterington}, David J.},
        title = "{Precise measurements of self-absorbed rising reverse shock emission from gamma-ray burst 221009A}",
      journal = {Nature Astronomy},
         year = 2023,
        month = aug,
       volume = {7},
        pages = {986-995},
          doi = {10.1038/s41550-023-01997-9}}

@ARTICLE{2021Galax...9...82G,
       author = {{Gill}, Ramandeep and {Kole}, Merlin and {Granot}, Jonathan},
        title = "{GRB Polarization: A Unique Probe of GRB Physics}",
      journal = {Galaxies},
         year = 2021,
        month = oct,
       volume = {9},
       number = {4},
          eid = {82},
        pages = {82},
          doi = {10.3390/galaxies9040082},
archivePrefix = {arXiv},
       eprint = {2109.03286},
 primaryClass = {astro-ph.HE},
       adsurl = {https://ui.adsabs.harvard.edu/abs/2021Galax...9...82G}
}

@ARTICLE{2019ApJ...884L..58U,
       author = {{Urata}, Yuji and {Toma}, Kenji and {Huang}, Kuiyun and {Asada}, Keiichi and {Nagai}, Hiroshi and {Takahashi}, Satoko and {Petitpas}, Glen and {Tashiro}, Makoto and {Yamaoka}, Kazutaka},
        title = "{First Detection of Radio Linear Polarization in a Gamma-Ray Burst Afterglow}",
      journal = {\apjl},
         year = 2019,
        month = oct,
       volume = {884},
       number = {2},
          eid = {L58},
        pages = {L58},
          doi = {10.3847/2041-8213/ab48f3},
archivePrefix = {arXiv},
       eprint = {1904.08111},
 primaryClass = {astro-ph.HE},
       adsurl = {https://ui.adsabs.harvard.edu/abs/2019ApJ...884L..58U}
}

@ARTICLE{2021ApJ...913...58S,
       author = {{Shimoda}, Jiro and {Toma}, Kenji},
        title = "{Multi-wave band Synchrotron Polarization of Gamma-Ray Burst Afterglows}",
      journal = {\apj},
         year = 2021,
        month = may,
       volume = {913},
       number = {1},
          eid = {58},
        pages = {58},
          doi = {10.3847/1538-4357/abf2c2},
archivePrefix = {arXiv},
       eprint = {2005.03710},
 primaryClass = {astro-ph.HE},
       adsurl = {https://ui.adsabs.harvard.edu/abs/2021ApJ...913...58S}
}

@ARTICLE{2018ApJ...862...94L,
       author = {{Laskar}, Tanmoy and {Alexander}, Kate D. and {Berger}, Edo and {Guidorzi}, Cristiano and {Margutti}, Raffaella and {Fong}, Wen-fai and {Kilpatrick}, Charles D. and {Milne}, Peter and {Drout}, Maria R. and {Mundell}, C.~G. and {Kobayashi}, Shiho and {Lunnan}, Ragnhild and {Barniol Duran}, Rodolfo and {Menten}, Karl M. and {Ioka}, Kunihito and {Williams}, Peter K.~G.},
        title = "{First ALMA Light Curve Constrains Refreshed Reverse Shocks and Jet Magnetization in GRB 161219B}",
      journal = {\apj},
         year = 2018,
        month = aug,
       volume = {862},
       number = {2},
          eid = {94},
        pages = {94},
          doi = {10.3847/1538-4357/aacbcc},
archivePrefix = {arXiv},
       eprint = {1808.09476},
 primaryClass = {astro-ph.HE},
       adsurl = {https://ui.adsabs.harvard.edu/abs/2018ApJ...862...94L}
}

@ARTICLE{2026ApJ..1005L..31K,
       author = {{Keating}, Garrett K. and {Laskar}, Tanmoy and {Ho}, Anna Y.~Q. and {Blanchard}, Peter K. and {Alexander}, Kate D. and {Berger}, Edo and {Gurwell}, Mark and {Eftekhari}, Tarraneh and {Xu}, Chloe T. and {Lovell}, Joshua Bennett and {Rao}, Ramprasad and {Williams}, Peter K.~G.},
        title = "{Rapid-response 1.3 mm Observations of GRB 260127A with the Submillimeter Array}",
      journal = {\apjl},
         year = 2026,
        month = jul,
       volume = {1005},
       number = {1},
          eid = {L31},
        pages = {L31},
          doi = {10.3847/2041-8213/ae5b7b},
archivePrefix = {arXiv},
       eprint = {2604.14297},
 primaryClass = {astro-ph.HE},
       adsurl = {https://ui.adsabs.harvard.edu/abs/2026ApJ..1005L..31K}
}

@ARTICLE{2012A&A...538A..44D,
       author = {{de Ugarte Postigo}, A. and {Lundgren}, A. and {Mart{\'\i}n}, S. and {Garcia-Appadoo}, D. and {de Gregorio Monsalvo}, I. and {Peck}, A. and {Micha{\l}owski}, M.~J. and {Th{\"o}ne}, C.~C. and {Campana}, S. and {Gorosabel}, J. and {Tanvir}, N.~R. and {Wiersema}, K. and {Castro-Tirado}, A.~J. and {Schulze}, S. and {De Breuck}, C. and {Petitpas}, G. and {Hjorth}, J. and {Jakobsson}, P. and {Covino}, S. and {Fynbo}, J.~P.~U. and {Winters}, J.~M. and {Bremer}, M. and {Levan}, A.~J. and {Llorente}, A. and {S{\'a}nchez-Ram{\'\i}rez}, R. and {Tello}, J.~C. and {Salvaterra}, R.},
        title = "{Pre-ALMA observations of GRBs in the mm/submm range}",
      journal = {\aap},
         year = 2012,
        month = feb,
       volume = {538},
          eid = {A44},
        pages = {A44},
          doi = {10.1051/0004-6361/201117848},
archivePrefix = {arXiv},
       eprint = {1108.1797},
 primaryClass = {astro-ph.CO},
       adsurl = {https://ui.adsabs.harvard.edu/abs/2012A&A...538A..44D}
}

@ARTICLE{bosch2026constraining,
       author = {{Bosch-Cabot}, Pau and {Tetarenko}, Alexandra J. and {Rosolowsky}, Erik and {Carotenuto}, Francesco and {Miller-Jones}, James and {Russell}, David M. and {Corbel}, St{\'e}phane and {Russell}, Thomas D. and {Sivakoff}, Gregory R.},
        title = "{Constraining the Jet Energetics of the Transient X-Ray Binaries MAXI J1348{\ensuremath{-}}630 and MAXI J1820+070 through Calorimetry}",
      journal = {\apj},
         year = 2026,
        month = jan,
       volume = {997},
       number = {1},
          eid = {64},
        pages = {64},
          doi = {10.3847/1538-4357/ae20f2},
archivePrefix = {arXiv},
       eprint = {2511.14941},
 primaryClass = {astro-ph.HE},
       adsurl = {https://ui.adsabs.harvard.edu/abs/2026ApJ...997...64B}
}

@ARTICLE{Yang2016,
       author = {{Yang}, J. and {Paragi}, Z. and {van der Horst}, A.~J. and {Gurvits}, L.~I. and {Campbell}, R.~M. and {Giannios}, D. and {An}, T. and {Komossa}, S.},
        title = "{No apparent superluminal motion in the first-known jetted tidal disruption event Swift J1644+5734}",
      journal = {\mnras},
         year = 2016,
        month = oct,
       volume = {462},
       number = {1},
        pages = {L66-L70},
          doi = {10.1093/mnrasl/slw107},
archivePrefix = {arXiv},
       eprint = {1605.06461},
 primaryClass = {astro-ph.HE},
       adsurl = {https://ui.adsabs.harvard.edu/abs/2016MNRAS.462L..66Y}
}

@ARTICLE{Zauderer2011,
       author = {{Zauderer}, B.~A. and {Berger}, E. and {Soderberg}, A.~M. and {Loeb}, A. and {Narayan}, R. and {Frail}, D.~A. and {Petitpas}, G.~R. and {Brunthaler}, A. and {Chornock}, R. and {Carpenter}, J.~M. and {Pooley}, G.~G. and {Mooley}, K. and {Kulkarni}, S.~R. and {Margutti}, R. and {Fox}, D.~B. and {Nakar}, E. and {Patel}, N.~A. and {Volgenau}, N.~H. and {Culverhouse}, T.~L. and {Bietenholz}, M.~F. and {Rupen}, M.~P. and {Max-Moerbeck}, W. and {Readhead}, A.~C.~S. and {Richards}, J. and {Shepherd}, M. and {Storm}, S. and {Hull}, C.~L.~H.},
        title = "{Birth of a relativistic outflow in the unusual {\ensuremath{\gamma}}-ray transient Swift J164449.3+573451}",
      journal = {\nat},
         year = 2011,
        month = aug,
       volume = {476},
       number = {7361},
        pages = {425-428},
          doi = {10.1038/nature10366},
archivePrefix = {arXiv},
       eprint = {1106.3568},
 primaryClass = {astro-ph.HE},
       adsurl = {https://ui.adsabs.harvard.edu/abs/2011Natur.476..425Z}
}

@ARTICLE{Zauderer2013,
       author = {{Zauderer}, B.~A. and {Berger}, E. and {Margutti}, R. and {Pooley}, G.~G. and {Sari}, R. and {Soderberg}, A.~M. and {Brunthaler}, A. and {Bietenholz}, M.~F.},
        title = "{Radio Monitoring of the Tidal Disruption Event Swift J164449.3+573451. II. The Relativistic Jet Shuts Off and a Transition to Forward Shock X-Ray/Radio Emission}",
      journal = {\apj},
         year = 2013,
        month = apr,
       volume = {767},
       number = {2},
          eid = {152},
        pages = {152},
          doi = {10.1088/0004-637X/767/2/152},
archivePrefix = {arXiv},
       eprint = {1212.1173},
 primaryClass = {astro-ph.HE},
       adsurl = {https://ui.adsabs.harvard.edu/abs/2013ApJ...767..152Z}
}

@ARTICLE{Berger2012,
       author = {{Berger}, E. and {Zauderer}, A. and {Pooley}, G.~G. and {Soderberg}, A.~M. and {Sari}, R. and {Brunthaler}, A. and {Bietenholz}, M.~F.},
        title = "{Radio Monitoring of the Tidal Disruption Event Swift J164449.3+573451. I. Jet Energetics and the Pristine Parsec-scale Environment of a Supermassive Black Hole}",
      journal = {\apj},
         year = 2012,
        month = mar,
       volume = {748},
       number = {1},
          eid = {36},
        pages = {36},
          doi = {10.1088/0004-637X/748/1/36},
archivePrefix = {arXiv},
       eprint = {1112.1697},
 primaryClass = {astro-ph.HE},
       adsurl = {https://ui.adsabs.harvard.edu/abs/2012ApJ...748...36B}
}

@ARTICLE{Metzger2012,
       author = {{Metzger}, Brian D. and {Giannios}, Dimitrios and {Mimica}, Petar},
        title = "{Afterglow model for the radio emission from the jetted tidal disruption candidate Swift J1644+57}",
      journal = {\mnras},
         year = 2012,
        month = mar,
       volume = {420},
       number = {4},
        pages = {3528-3537},
          doi = {10.1111/j.1365-2966.2011.20273.x},
archivePrefix = {arXiv},
       eprint = {1110.1111},
 primaryClass = {astro-ph.HE},
       adsurl = {https://ui.adsabs.harvard.edu/abs/2012MNRAS.420.3528M}
}

@ARTICLE{Eftekhari2018,
       author = {{Eftekhari}, T. and {Berger}, E. and {Zauderer}, B.~A. and {Margutti}, R. and {Alexander}, K.~D.},
        title = "{Radio Monitoring of the Tidal Disruption Event Swift J164449.3+573451. III. Late-time Jet Energetics and a Deviation from Equipartition}",
      journal = {\apj},
         year = 2018,
        month = feb,
       volume = {854},
       number = {2},
          eid = {86},
        pages = {86},
          doi = {10.3847/1538-4357/aaa8e0},
archivePrefix = {arXiv},
       eprint = {1710.07289},
 primaryClass = {astro-ph.HE},
       adsurl = {https://ui.adsabs.harvard.edu/abs/2018ApJ...854...86E}
}

@ARTICLE{Cendes2021,
       author = {{Cendes}, Y. and {Eftekhari}, T. and {Berger}, E. and {Polisensky}, E.},
        title = "{Radio Monitoring of the Tidal Disruption Event Swift J164449.3+573451. IV. Continued Fading and Non-relativistic Expansion}",
      journal = {\apj},
         year = 2021,
        month = feb,
       volume = {908},
       number = {2},
          eid = {125},
        pages = {125},
          doi = {10.3847/1538-4357/abd323},
archivePrefix = {arXiv},
       eprint = {2011.00074},
 primaryClass = {astro-ph.HE},
       adsurl = {https://ui.adsabs.harvard.edu/abs/2021ApJ...908..125C}
}

@ARTICLE{Cendes2022,
       author = {{Cendes}, Y. and {Berger}, E. and {Alexander}, K.~D. and {Gomez}, S. and {Hajela}, A. and {Chornock}, R. and {Laskar}, T. and {Margutti}, R. and {Metzger}, B. and {Bietenholz}, M.~F. and {Brethauer}, D. and {Wieringa}, M.~H.},
        title = "{A Mildly Relativistic Outflow Launched Two Years after Disruption in Tidal Disruption Event AT2018hyz}",
      journal = {\apj},
         year = 2022,
        month = oct,
       volume = {938},
       number = {1},
          eid = {28},
        pages = {28},
          doi = {10.3847/1538-4357/ac88d0},
archivePrefix = {arXiv},
       eprint = {2206.14297},
 primaryClass = {astro-ph.HE},
       adsurl = {https://ui.adsabs.harvard.edu/abs/2022ApJ...938...28C}
}

@ARTICLE{Cendes2024,
       author = {{Cendes}, Y. and {Berger}, E. and {Alexander}, K.~D. and {Chornock}, R. and {Margutti}, R. and {Metzger}, B. and {Wieringa}, M.~H. and {Bietenholz}, M.~F. and {Hajela}, A. and {Laskar}, T. and {Stroh}, M.~C. and {Terreran}, G.},
        title = "{Ubiquitous Late Radio Emission from Tidal Disruption Events}",
      journal = {\apj},
         year = 2024,
        month = aug,
       volume = {971},
       number = {2},
          eid = {185},
        pages = {185},
          doi = {10.3847/1538-4357/ad5541},
archivePrefix = {arXiv},
       eprint = {2308.13595},
 primaryClass = {astro-ph.HE},
       adsurl = {https://ui.adsabs.harvard.edu/abs/2024ApJ...971..185C}
}

@ARTICLE{BarniolDuran2013,
       author = {{Barniol Duran}, Rodolfo and {Nakar}, Ehud and {Piran}, Tsvi},
        title = "{Radius Constraints and Minimal Equipartition Energy of Relativistically Moving Synchrotron Sources}",
      journal = {\apj},
         year = 2013,
        month = jul,
       volume = {772},
       number = {1},
          eid = {78},
        pages = {78},
          doi = {10.1088/0004-637X/772/1/78},
archivePrefix = {arXiv},
       eprint = {1301.6759},
 primaryClass = {astro-ph.HE},
       adsurl = {https://ui.adsabs.harvard.edu/abs/2013ApJ...772...78B}
}

@ARTICLE{Beniamini2023,
       author = {{Beniamini}, Paz and {Piran}, Tsvi and {Matsumoto}, Tatsuya},
        title = "{Swift J1644+57 as an off-axis Jet}",
      journal = {\mnras},
         year = 2023,
        month = sep,
       volume = {524},
       number = {1},
        pages = {1386-1395},
          doi = {10.1093/mnras/stad1950},
archivePrefix = {arXiv},
       eprint = {2305.06370},
 primaryClass = {astro-ph.HE},
       adsurl = {https://ui.adsabs.harvard.edu/abs/2023MNRAS.524.1386B}
}

@ARTICLE{Horesh2021,
       author = {{Horesh}, A. and {Cenko}, S.~B. and {Arcavi}, I.},
        title = "{Delayed radio flares from a tidal disruption event}",
      journal = {Nature Astronomy},
         year = 2021,
        month = may,
       volume = {5},
        pages = {491-497},
          doi = {10.1038/s41550-021-01300-8},
archivePrefix = {arXiv},
       eprint = {2102.11290},
 primaryClass = {astro-ph.HE},
       adsurl = {https://ui.adsabs.harvard.edu/abs/2021NatAs...5..491H}
}

@ARTICLE{Bally2016,
       author = {{Bally}, John},
        title = "{Protostellar Outflows}",
      journal = {\araa},
         year = 2016,
        month = sep,
       volume = {54},
        pages = {491-528},
          doi = {10.1146/annurev-astro-081915-023341},
       adsurl = {https://ui.adsabs.harvard.edu/abs/2016ARA&A..54..491B}
}
\end{document}